\documentclass[twocolumn,preprintnumbers,amsmath,amssymb]{revtex4}

\usepackage{graphicx}
\usepackage{dcolumn}
\usepackage{bm}
\usepackage{color}

\begin{document}


\title{\color{black} \color{black} Two \color{black} phase transitions induced by a magnetic field in graphite}
\author{Beno\^{\i}t Fauqu\'e$^{1,\footnote{benoit.fauque@espci.fr}}$, David LeBoeuf$^2$, Baptiste Vignolle$^2$, Marc Nardone$^2$, Cyril Proust$^2$ and Kamran Behnia$^1$}%

\affiliation{
$^1$ LPEM (UPMC-CNRS), Ecole Sup\'erieure de Physique et de Chimie Industrielles, 75005 Paris, France\\
$^2$ Laboratoire National des Champs Magn\'etiques Intenses (CNRS-INSA-UJF-UPS), 31400 Toulouse, France\\
}


\begin{abstract}
Different instabilities have been speculated for a three-dimensional electron gas confined to its lowest Landau level.  The phase transition induced in graphite by a strong magnetic field, and believed to be a Charge Density Wave (CDW), is the only experimentally established case of such instabilities. Studying the magnetoresistance in graphite for the first time up to 80 T, we find that the magnetic field induces two successive phase transitions, consisting of two distinct ordered states each restricted to a finite field window. In both states, an energy gap opens up in the out-of-plane conductivity and coexists with an unexpected in-plane metallicity for a fully gap bulk system. Such peculiar metallicity may arise as a consequence of edge-state transport expected to develop in presence of a bulk gap.
\end{abstract}

\pacs{Valid PACS appear here}
\maketitle
\begin{figure}[htbp]
\begin{center}
\includegraphics[angle=0,width=7.0cm]{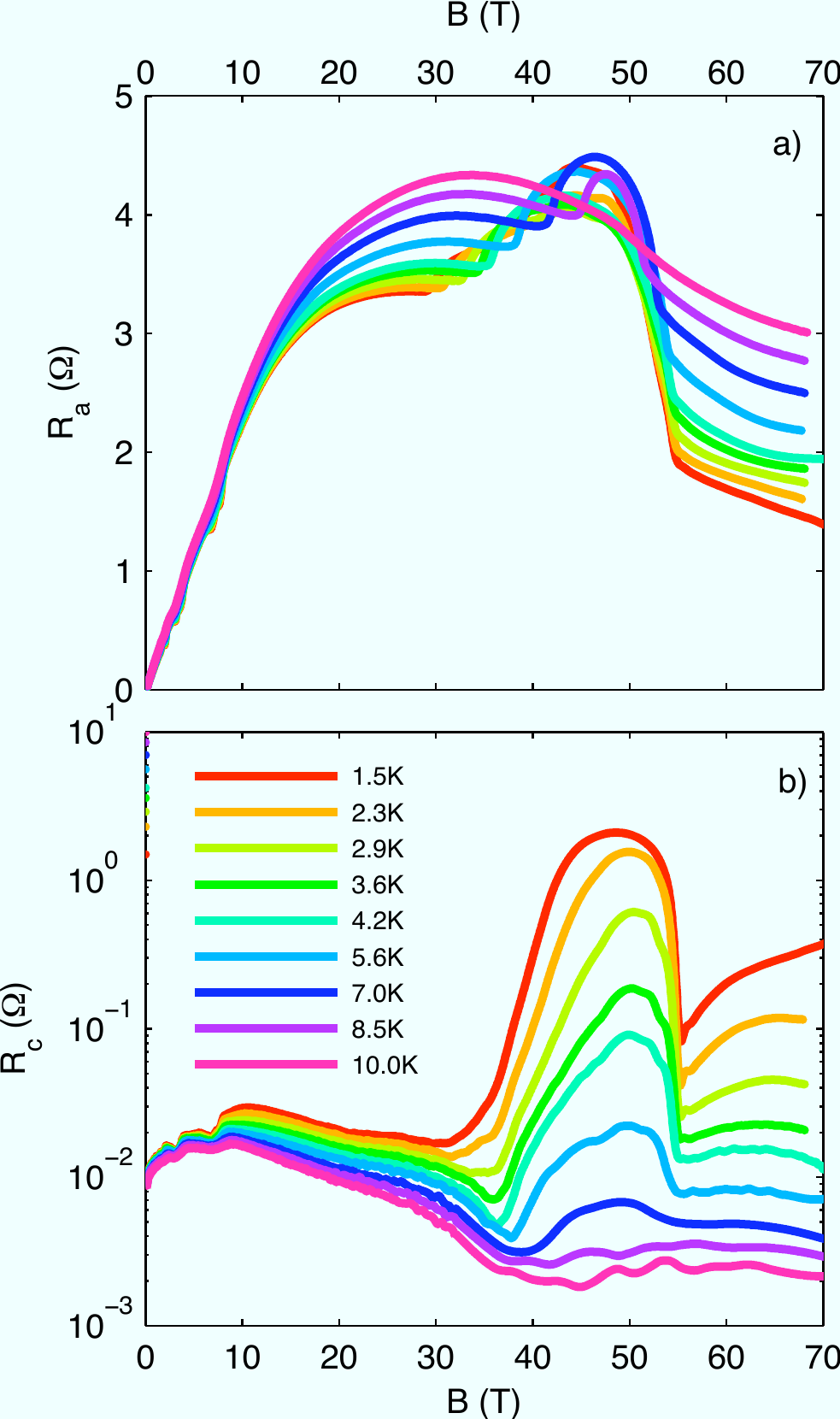}
\caption{  a) In-plane magnetoresistance of kish graphite (sample K$_1$) for different temperatures. b) Out-of-plane magnetoresistance of sample (K$_2$)  for various temperature. The magnetic field is oriented along the c-axis.  Note the semi-logarithmic scale of the lower panel.}
\label{Fig1}
\end{center}
\end{figure}

 \color{black}
Graphite is a compensated semi-metal with a tiny three-dimensional Fermi surface and is described by the Slonczewski-Weiss-McClure (SWM) band model \cite{SW1958,McClure1957}.  A modest  magnetic field of  7.5 T perpendicular to the graphene layers confines electrons and holes  to their lowest Landau levels (LLs). It is therefore an ideal candidate to explore the nature of the electronic ground state of a three-dimensional electron gas pushed beyond the so-called quantum limit. In this regime  \cite{zhu2010}, the electronic spectrum becomes analogue to a one-dimensional system with a variety of possible instabilities\cite{MacDonald87,Halperin1987,Biagini2001}. \color{black}
In the early eighties, a phase transition in this system was discovered by Tanuma and co-workers who reported a sharp increase in the in-plane magnetoresistance of graphite at B$\approx$25 T \cite{tanuma1981}. Numerous experimental studies followed \cite{iye1982,iye1984,ochimizu1992,yaguchi1993,yaguchi1998,yaguchi2001,Fauque2011} and confirmed the existence of a field-induced many-body state beyond a temperature-dependent critical magnetic field. Soon after the initial experimental discovery, Yoshioka and Fukuyama (YF) \cite{Yoshioka1981} ascribed this instability to a charge-density-wave (CDW). Such an instability is favored by one-dimensionality, because of the availability of a suitable (2k$_F$) nesting vector. In the original YF scenario,  the CDW in adjacent valleys are out of phase, creating a valley density wave state \cite{Tesanovic1987}, in order to minimize Coloumb energy.  In 1998, by further increasing the magnetic field, Yaguchi and Singleton found that the field-induced state is eventually destroyed beyond 53 T \cite{yaguchi1998}. This destruction was attributed to the depopulation of a Landau sub-level within the framework of YF scenario. The possible multiplicity of induced phases and their signatures in the in-plane and out of-plane charge transport remain open questions (for a review see \cite{Yaguchi2009}).  Therefore, despite a large body of research on graphite at high magnetic field, the nature of its electronic ground state is not settled. In addition, these investigations can provide interesting input to the debate on the importance of many-body physics and its evolution with dimensionality in graphene \cite{Kotov2013}.



In this letter, we extend the previous measurements up to a magnetic field as strong as 80 T and focus on the contrast between in-plane and out-of-plane resistivity. Our results lead to a significant revision of the experimental picture. As the magnetic field is swept, an electronic state is indeed first induced and then destroyed by magnetic field. But this  state is immediately followed by another and almost identical one, which is also first induced and then destroyed by the increasing magnetic field. We resolve an activated behavior in the temperature dependence of the out-of-plane conductivity in both of these ordered states, which allows us to quantify the magnitude of the out-of-plane charge gap for the first time. Interestingly, such a gap is absent in the in-plane transport. We argue that this may be the first observation of the three-dimensional counterpart to the two-dimensional edge states known to occur in quantum Hall systems

Standard four-probe resistivity measurements were performed on both kish and HOPG samples in pulsed magnetic fields at LNCMI-Toulouse. A detailed description of experimental procedure is given in Supplementary Information (S.I.) section A. Fig.\ref{Fig1} presents the data for in-plane and out-of plane magnetoresistance up to 70 T for different temperatures. As seen in Fig.\ref{Fig1}.a, as a consequence of high mobility, the  in-plane magnetoresistance  is very large. It saturates around B$\sim20$ T before becoming slightly negative. The onset of the transition to the field-induced state is marked by a sudden increase in resistance. This jump is followed by a drop corresponding to reentry to a low-resistive state at 53 T. Above 53 T, the magnetoresistivity continues to decrease smoothly without indicating any new field scale. These results are in good agreement with the previous data up to 54 T reported by Yaguchi and Singleton\cite{yaguchi1998}.

Magnetoresistance along $c$-axis, $R_{c}$, presented in Fig.\ref{Fig1}.b, shows several remarkable differences with  $R_{a}$ \color{black}( in plane magnetoresistance ). In all the experiments, the magnetic field is oriented along the c-axis. \color{black}. First, below the onset of transition, the out-of-plane magnetoresistance is much smaller than the in plane magnetoresistance. This is not surprising, since the Lorentz force does not affect the electron motion along the magnetic field\cite{Maslov2010} (See the S. I section B for a more detailed discussion of the low field magnetoresistance). A second difference between $R_{c}$ and $R_{a}$ is the amplitude of the variation caused by the phase transition. In the case of R$_{a}$, it is less than a factor of two, whereas in the case of R$_{c}$, it is \color{black} three orders of magnitude\color{black}. The remarkable sensitivity of the $c$-axis transport has been noticed previously\cite{yaguchi2001}, but never quantified. The third difference between $R_{c}$ and $R_{a}$ is their diverging behavior above 53 T. In this range, R$_{c}$ increases with magnetic field and this enhancement strongly depends on the temperature. On the other hand, R$_{a}$ decreases with increasing magnetic field or decreasing temperature.

\begin{figure}[htbp]
\begin{center}
\includegraphics[angle=0,width=6.5cm]{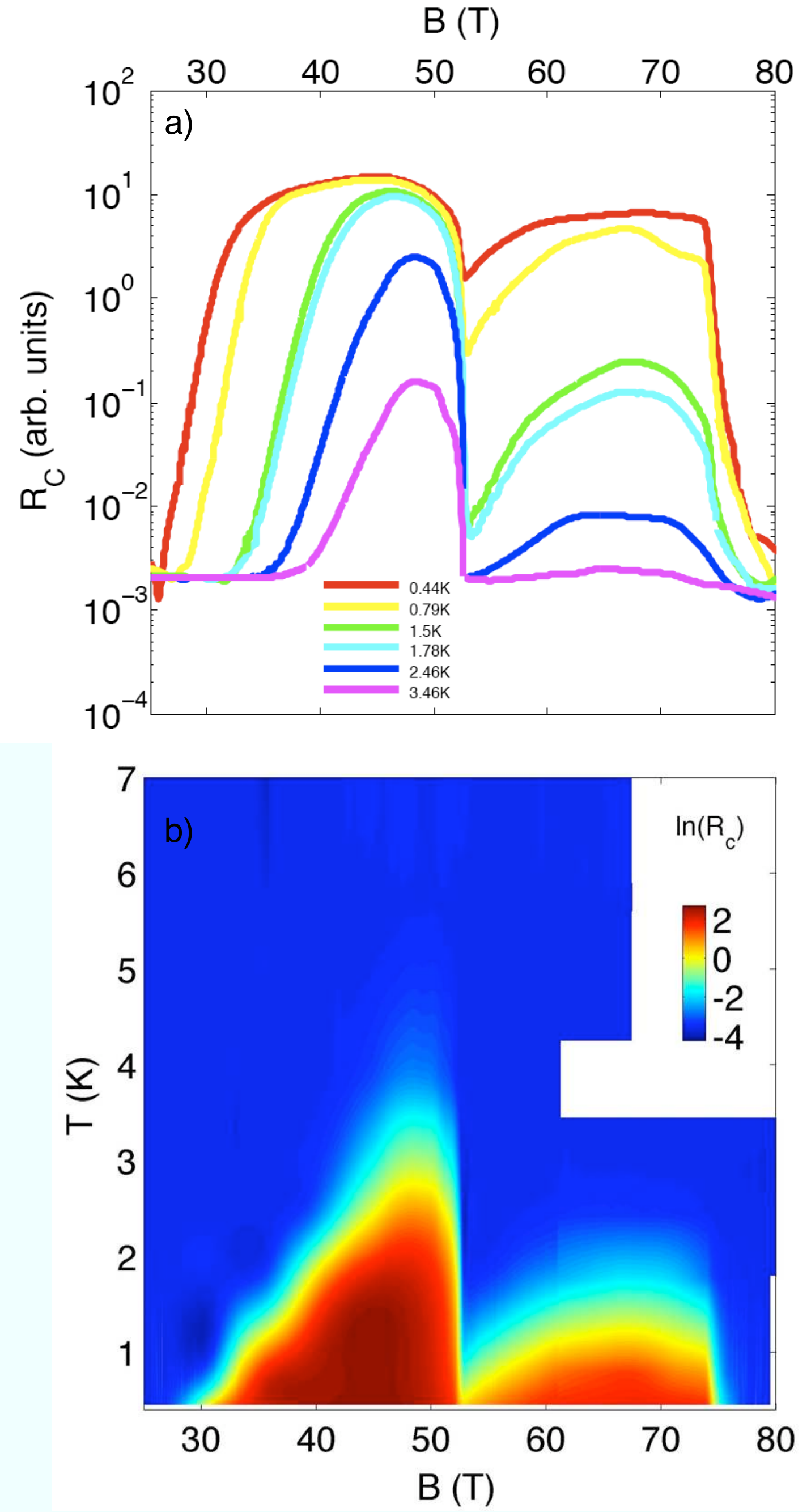}
\caption{  a) Out-of-plane magnetoresistance in semi-logarithmic scale of kish graphite (sample K$_3$)  up to 80 T for the lowest investigated temperatures. b) A logarithmic color map of R$_c$  ( sample K$_3$) in the plane of temperature and magnetic field.}
\label{Fig2}
\end{center}
\end{figure}

Before this work, the system was believed to reenter its low-field state above 53 T\cite{Yaguchi2009}. Therefore, the increase in R$_c$ above this field was unexpected. This observation motivated us to extend our investigation of R$_c$ to higher fields (80 T) and lower temperatures (0.44 K). The data are presented in Fig.\ref{Fig2}.a. As seen in the figure, above 53 T, R$_c$ increases, saturates and then steeply drops at 75 T to recover its magnitude before the onset of transition.  Thus, R$_c$ is enhanced by several orders of magnitude in two adjacent yet distinct field windows. The first extends between 30 T to 53 T and the second between 53 T and 75 T. In both  windows, the amplitude of R$_c$ becomes extremely temperature-dependent. Fig.\ref{Fig2}.b presents temperature and field dependence of R$_c$ in a logarithmic color map. A new dome clearly emerges above 53 T. This corresponds to the boundaries of a new electronic phase in the (temperature, field) plane. The magnetoresistance recovers its normal amplitude above 3.5 K. This is the highest critical temperature of this new state. The two field scales 53 T and 75 T are temperature-independent contrary to the onset of the first transition which increases with increasing temperature. The identification of an additional phase in the  phase diagram of graphite beyond the quantum limit is the first result of this work. Similar features are observed in  the $c$-axis resistance measurements of HOPG sample (see the SI section B). Thus, the existence of these field induced states is a universal property of graphitic systems.  

\begin{figure}[htbp]
\begin{center}
\includegraphics[angle=0,width=7.5cm]{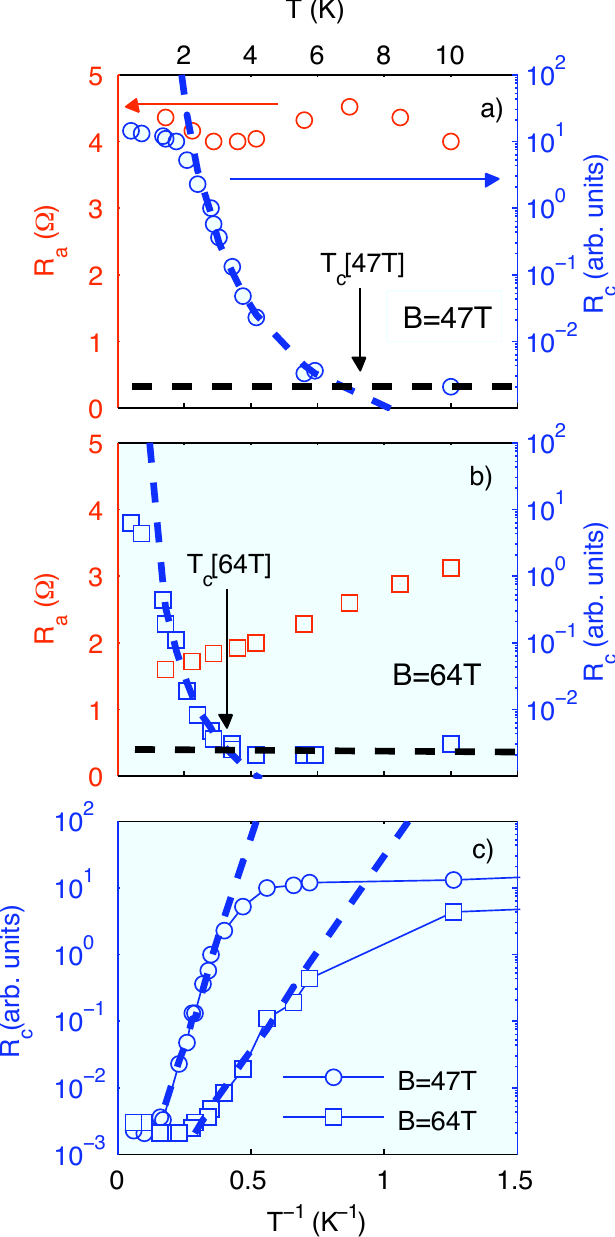}
\caption{  \color{black}Temperature dependence of the in-plane (red left axis)  and out-of-plane (blue right axis in log scale) magnetoresistance of kish graphite a) for B=47 T (circles points ) and b) for B=64 T (square points) for the two samples. Activated (temperature-independent) variations are represented by blue (black) dashed lines. The critical temperature of the field-induced transition is defined as the intercept of these two lines. c)  R$_c$  as a function  of T$^{-1}$ for  B=47 T (circles) and B=64 T (squares). Dashed lines are Arrhenius fits to R$_c$.\color{black}}
\label{Fig3}
\end{center}
\end{figure}

The second result is presented in Fig.\ref{Fig3}, which compares the temperature-dependence of the in-plane and out-of-plane resistivity in the two ordered states.  In  Fig.\ref{Fig3}.a) and b) (red left scale), data for R$_a$ are presented for  B=47 T and B=64 T. At each of these fields, upon cooling, the system enters deep inside one of the two ordered states (see Fig.\ref{Fig2}b). \color{black} In neither case does the in-plane resistivity display insulating behavior in the ordered state\color{black}. The temperature dependence of R$_c$ for the same fields is presented in Fig.\ref{Fig3}.a) and b) (blue right scale). For both fields, plotting R$_c$ as a function of T$^{-1}$ (Fig.\ref{Fig2}.c)) reveals an Arrhenius behavior upon the entry to the ordered state. In both cases, at low temperature, resistance deviates from the Arrhenius behavior and saturates to a finite value pointing to the survival of a residual conductivity along c-axis at low temperature. We will come back to the origin of this residual c-axis conductivity in the discussion below.

 \color{black} By fitting the data in the activated regime to R$_c\propto \exp(\frac{2\Delta}{k_B T})$ \color{black}, we found activation gaps to be $2 \Delta[47 T]=2.4meV$ and $2\Delta[64 T]=1.1meV$.  For each field, a critical temperature (T$_c$) for these transitions can be defined as the temperature at which the Arrhenius fit (blue dot lines on Fig.\ref{Fig3}.a) and b)) crosses the normal state resistance value (black dot lines on Fig.\ref{Fig3}.a) and b)). This yields T$_c$[47T]=7$\pm$0.5K and T$_c$[64T]=3.5$\pm$0.5K. Comparing the experimentally-resolved activation gap with the  critical temperature, we find a similar ratio for both states, $\frac{2\Delta}{k_B T_{c}}[47 T]=3.9$ and $\frac{2\Delta}{k_B T_{c}}[64 T]=3.6$. Thus, the second ordered state has a gap twice smaller and survives up to a temperature twice lower that the first ordered state.

\begin{figure}[htbp]
\begin{center}
\includegraphics[angle=0,width=8.25cm]{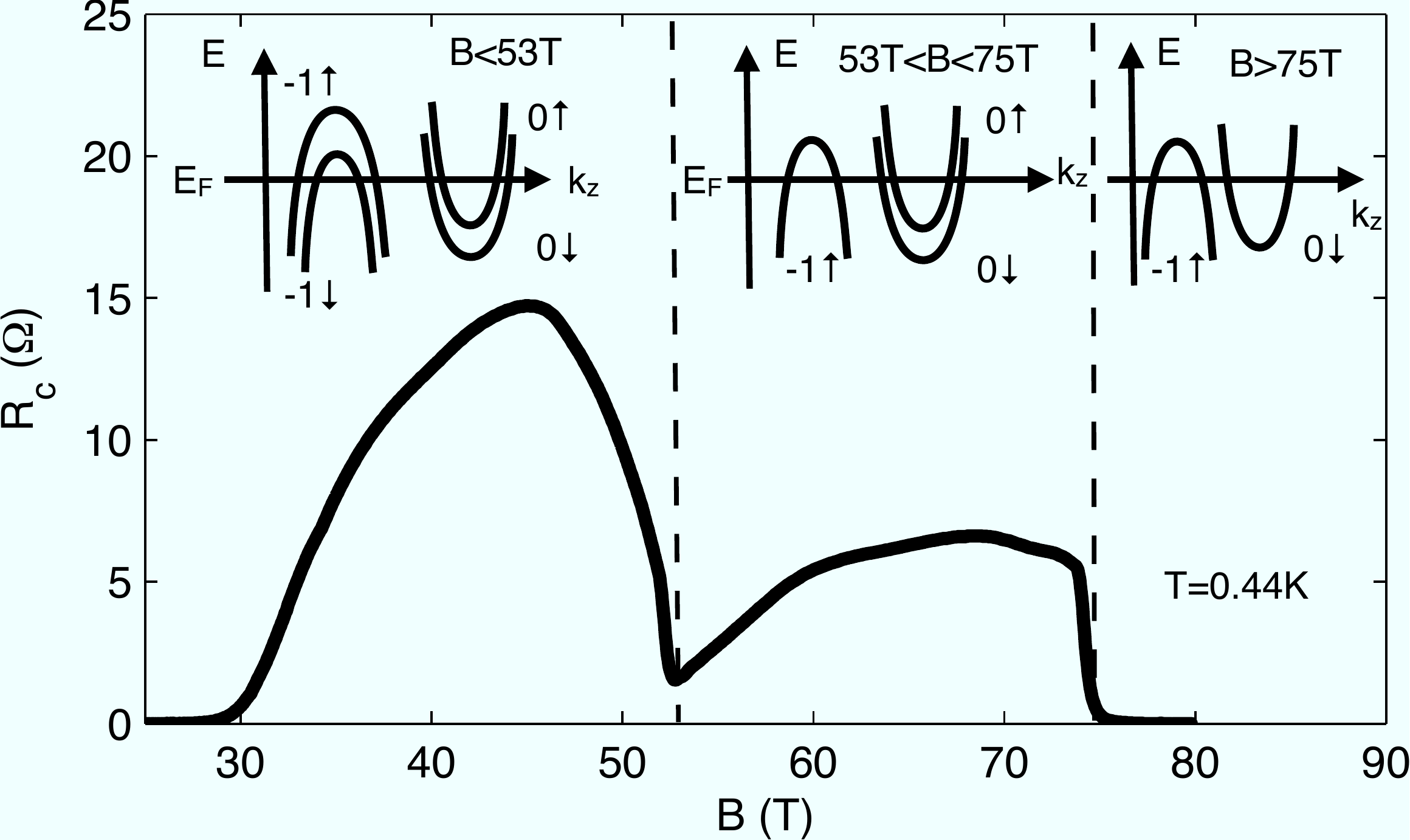}
\caption{   Field dependence of the out-of-plane resistance R$_c$ of kish graphite (sample K$_3$) at T=0.44K ( black line).  Sketch of occupied Landau sub-levels with increasing  magnetic field. When B$>$7.5 T, there are four occupied sub-levels. The onset of field-induced transition at 30 T, opens a gap in all these sub-levels. At  B=53 T,  one sub-level crosses the Fermi energy and the field-induced state is destroyed. For B$>$53 T, a second field-induced state emerges opening a gap in the three last occupied sub-levels. At B=75 T,  another sub-level becomes empty and the second field-induced state is destroyed. \color{black}}
\label{Fig4}
\end{center}
\end{figure}

These observations lead to a \color{black} revision \color{black} of our current understanding of the physics of the field-induced phase transition in graphite. Two possible configurations have been proposed for the density wave observed below 53 T. Either two charge density waves (per valley) formed in both (n=0,$\downarrow$) and (n=0,$\uparrow$) sub-levels of electrons (and respectively in both (n=-1,$\downarrow$) and (n=-1,$\uparrow$) sub-levels of holes)\cite{Yoshioka1981} or one spin density wave (per valley) in each electron and hole pocket\cite{Takahashi94}. Yaguchi and Singleton\cite{yaguchi1998} suggested that the two Landau sub-levels (n=0,$\uparrow$) (for the electrons) and (n=-1,$\downarrow$) (for the holes) are depopulated at 53 T and this destroys the CDW state. In this case, above this field, the two available Landau levels would be (n=0,$\downarrow$) and (n=-1,$\uparrow$). Following this picture, one should assume that the system undergoes a second density-wave instability along the c-axis at 53 T. In this case, graphite would host a cascade of density-wave transitions, a scenario reminiscent of what has been proposed in the case of the quasi-one dimensional organic conductor (Per)$_2$Pt(mnt)$_2$ \cite{Graf2004}.

However, this interpretation is challenged by the small and almost temperature-independent resistivity detected above 75 T. In the picture drawn above, when the magnetic field exceeds 53 T, only the two last spin-polarized Landau sub-levels would remain occupied. This reduces the number of possible configurations for any nesting scenario. If two sub-levels empty at 53 T and two others at 75 T, one would expect an ultimate high-field semiconducting state in contrast to what is observed experimentally. Persistence of a sizable conductivity above 75 T is not compatible with a total depopulation of all Landau sub-levels above this field.

A tempting solution  would be to revise this scenario and assume that at B=53 T only one sub-level (instead of two) depopulates (see the sketch in Fig.\ref{Fig4}). This could be either of the two (n=0,$\uparrow$) or (n=-1,$\downarrow$) Landau sub-levels. In this case, for B$>$ 53 T, there are three occupied sub-levels and a second gapped state. At B=75 T, the other sub-level is depopulated and the ordered state is destroyed. In this case, beyond 75 T, the ultimate electron and hole Landau sub-levels will remain full, in agreement with the finite conductivity observed by our experiment. In the SWM model, which is in very good agreement with the experimental low-field Landau spectrum \cite{zhu2010,Schneider2009}, the hole-like Fermi surface has a slightly smaller cross section than the electron-like one. Therefore, the (n=-1,$\downarrow$) sub-level is expected to become empty at a lower magnetic field than the (n=0,$\uparrow$) sub-level (as sketched in Fig.\ref{Fig4}). However, according to Takada and Goto\cite{Takada1999}, electron correlations modify the SWM spectrum at high magnetic field. Thus, the identity of the occupied sub-level between 53 T and 75 T remains an open question. This sub-level plays a crucial role in any nesting scenario, since both phase transitions occur when it is occupied and the order is destroyed as soon as it becomes empty.

Our most striking result is the observed difference between R$_a$ and R$_c$. To the best of our knowledge, the field-induced state in graphite is the only case for a 1D system in which activated conductivity along one axis coexists with metallic conductivity perpendicular to it. This dichotomy contrasts with what has been reported in other 1D density-wave systems. Bechgaard salts are a well-documented family of quasi-one dimensional conductors hosting a density-wave transition as a consequence of nesting. In the spin-density-wave of (TMTSF)$_2$PF$_6$, all three components of resistivity display an activated behavior with a gap of similar amplitude\cite{Dressel2005}. This is not surprising, since such a transition opens a gap in the Density Of States (DOS), which affects the conductivity along all orientations. 

 In our case,  the electronic spectrum of graphite is quantized by the magnetic field\color{black} . The energy distance between two Landau sub-levels is at least the Zeeman energy,  $g \mu_B B$ (where g=2.5 \cite{Maude2010}),  as large as 8 meV at 53 T. A three-dimensional solid keeps its continuous spectrum, in spite of such a gap, thanks to z-axis dispersion. However, as soon as the field-induced order opens a new gap, the spectrum of the bulk electrons is no longer continuous and they are not expected to display metallicity. On the other hand, the situation is different on the edge of the sample. One of the two gaps, or both of them, may close within a depth of magnetic length ($l_B=\sqrt(\frac{\hbar}{eB}) \sim 35 \AA$  at 53 T).

The existence of edge states for a 3D electron gas system beyond the quantum limit has been the subject of several theoretical studies\cite{Halperin1987,Bernevig2007,Burnell2009}. \color{black}As first pointed by Halperin\cite{Halperin1987}, quantum Hall effect can occur in three dimensions if the Fermi level of the bulk happens to lie inside an energy gap. A bulk density wave state along the $c$-axis would naturally fulfill such a condition. A theory for quantum Hall effect in graphite\cite{Bernevig2007} predicted the appearance of chiral surface states\cite{Balents1996} characterized by a ballistic in-plane longitudinal response. Such states have been observed in the multilayer semiconductor GaAs/Al$_{0.1}$Ga$_{0.9}$As by Druist \emph{et al.}\cite{Druist1998} who also found that these states have a finite vertical conductance. 

The order of magnitude of the residual conductivity measured here is a clue to its origin. Note that in-plane and out-of-plane charge flows mix up in graphite as a consequence of stacking defaults. Therefore, the measured intralayer and interlayer resistivities are to be treated cautiously and only their order of magnitude remains reliable. As seen in Fig.\ref{Fig3}.b), the residual in-plane conductivity is $\approx$1$\Omega^{-1}$ at 64 T. Since the thickness of our sample is 50$\mu m$ (150000 graphene layers), this corresponds to a conductivity of 0.15$\sigma_0$  per graphene layer ($\sigma_0$=$\frac{e^2}{h}$ is the quantum of conductance). Moreover, and as mentioned above (see Fig. \ref{Fig3}.c)),  a residual out-of plane conductivity survives in both ordered states at low temperature.  A possible source of the latter are impurity states, known to lead to a deviation from activated behavior at low temperatures\cite{Gruner1994}.  Another possibility is that this c-axis residual conductivity is the vertical conductivity of the edge states. Intriguingly, the ratio of $\frac{\rho_c}{\rho_{a}}$ at high magnetic field recovers its value at zero magnetic field (see the SI section C).

\color{black}

In summary, by investigating magnetoresistance of graphite up to 80T, we found a new electronic phase above 53 T. There are two distinct  states induced by the magnetic field. In both of them, the system is an insulator along the $c$-axis and metallic perpendicular to that axis. Such a dichotomy is surprising for a system with a full gap in the energy spectrum and may be explained by the presence of edge states. Therefore, the unexpectedly rich high-field phase diagram of graphite provides a novel perspective on the long-standing and unsolved nature of the electronic ground state of tridimensional electron gas confined to its lowest Landau level. 

 We thank J. Alicea, A. Bangura, M. Goerbig, D. K. Maude, D. Maslov, G. Montambaux, V. Oganesyan,Y. Takada and Y. Yakovenko for stimulating discussions and J. B\'eard, A. Zitouni for technical support during the 80 T experiment. This work is supported by the Agence Nationale de Recherche as a part of SUPERFIELD project, by a grant attributed by the Ile de France regional council and by EuroMagNET II under the EU contract number 228043.


\end{document}